\documentclass[a4paper,11pt]{article}

\pdfoutput=1 

\usepackage[T1]{fontenc} 

\usepackage[]{caption}
\usepackage{color,xcolor,graphicx,amsmath, amssymb,mathtools
, physics,ulem}

\usepackage{upgreek}			
\usepackage{bm}					
\newcommand{\g}{g_\star}			      		
\usepackage{booktabs} 

\newcommand{\be}{\begin{eqnarray}}
\newcommand{\ee}{\end{eqnarray}}
\newcommand{\bea}{\begin{eqnarray}}
\newcommand{\eea}{\end{eqnarray}}
\newcommand{\bi}{\bibitem}
\newcommand{\ct}{\cite}
\newcommand{\lt}{\left}
\newcommand{\rt}{\right}

\newcommand{\ba}{\begin{array}}
\newcommand{\ea}{\end{array}}
\newcommand{\bd}{\begin{displaymath}}
\newcommand{\ed}{\end{displaymath}}
\newcommand{\beq}{\begin{equation}}
\newcommand{\eeq}{\end{equation}}

\newcommand{\nn}{\nonumber}

\def\it{ {\em i.e.,\ }}

\def\g{\gamma}

\def\m{\mu}
\def\n{\nu}

\def\q2 {q^2}

\def\bt{\begin{table}}
\def\et{\end{table}}

\definecolor{shilamagenta}{rgb}{0.8, 0.0, 0.8}
\newcommand{\tcmag}{\textcolor{shilamagenta}}

\newcommand{\shilacred}{\color{red!70!green}}
\definecolor{shilagreen}{rgb}{0.0, 0.5, 0.0}
\definecolor{shilacyan}{rgb}{0.0, 0.58, 0.71}

\definecolor{midnightblue}{rgb}{0.1, 0.1, 0.44}

\usepackage{hyperref}

\hypersetup{citecolor=blue,        
    linkcolor=red,  
    filecolor=magenta,      
    urlcolor=shilacyan  
}

\usepackage{cleveref}
\usepackage{slashed}

\begin{document}

\title{\bf {\color{red} Entropy production due to electroweak phase transition in the framework of two Higgs doublet model}
}

\author{Arnab~Chaudhuri$^{a}$\footnote{{\bf e-mail}: \href{mailto:arnabchaudhuri.7@gmail.com}{arnabchaudhuri.7@gmail.com}},
Maxim Yu. Khlopov$^{b}$\footnote{{\bf e-mail:} \href{mailto:khlopov@apc.in2p3.fr}{khlopov@apc.in2p3.fr}},
\\
$^a$ \small{ Novosibirsk State University} \\
\small{ Pirogova ul., 2, 630090 Novosibirsk, Russia}\\
$^b$ \small{Institute of Physics, Southern Federal University}\\
\small{Stachki 194 Rostov on Don 344090, Russia}\\
\small{and Université de Paris, CNRS, Astroparticule et Cosmologie, F-75013 Paris, France}\\
\small{and National Research Nuclear University "MEPHI", Moscow 115409, Russia}\\
}

\date{}
\maketitle

\begin{abstract}
We revisit the possibility of first order electroweak phase transition~(EWPT) in one of the simplest extensions of the Standard Model (SM) scalar sector, namely the two-Higgs-doublet model~(2HDM). We take into account the ensuing constraints from the electroweak precision tests, Higgs signal strengths, and the recent LHC bounds from direct scalar searches. By studying the vacuum transition in 2HDM, we discuss in detail the entropy released in the first order EWPT in various parameter planes of 2HDM. 
\end{abstract}

\section{Introduction}

It is a well-established fact that EWPT is either a second order or a smooth crossover in the SM 
of particle physics. So is the fact that the entropy density in the early universe plasma is conserved in the course
of the cosmological expansion if the plasma is in thermal equilibrium state with negligible chemical potential of every species ~\cite{Gorbunov:2011zzc, Bambi:2015mba}. The entropy conservation law is given by
\be \label{ent-consv}
s=\frac{P+\rho}{T}a^3=const.
\ee
where $a(t)$ is the scale factor, $T(t)$ is the temperature of the fluid (or plasma), $\rho$ and $P$ are the energy density and pressure of the plasma respectively.

In the early universe, the state of matter is quite close to the equilibrium as the reaction rate  $\Gamma \sim n \sigma v$ is much faster than the cosmological expansion rate, i.e., the Hubble parameter $H=\Dot{a}/a\propto T^2/m_{Pl}$. The equilibrium condition $\Gamma > H$ is always satisfied for at temperature $T< \alpha \, m_{\rm Pl}$. Here $\alpha$ is the coupling constant of the particle interaction of the order of $\sim 10^{-2}$ and $m_{\rm Pl}$ is the Planck Mass. Due to the large value of $m_{\rm Pl}$, thermal equilibrium exists in most of the history of the universe, if $\alpha$ is $<<1$.

As mentioned above, during thermal equilibrium, the entropy density in the comoving volume is conserved. But there are scenarios where the entropy density is not conserved.
For example, if the universe at a certain stage was dominated by primordial blackholes~\cite{Dolgov:2000ht}, 
the entropy production can be very high, high enough to delete the pre-existing baryon asymmetry~\cite{Chaudhuri:2020wjo}. In the context of the modern cosmological paradigm of inflationary Universe with baryosynthesis and dark matter/energy, physics beyond the Standard model (BSM) underlying these necessary elements of the modern cosmological model can provide many examples of various mechanisms of high entropy production (see e.g. \cite{khlopovPPNP} for review and references). Taking apart the wide range of various possibilities we consider here the problem of entropy production by minimal extension of SM and start the discussion from the SM predictions for the cosmological entropy production.

A large entropy production could take place during QCD phase transition at $T \sim 100-200$MeV. But due to strong technical issues, QCD phase transition in early universe cosmology is not known in details. For reference, please see~\cite{Schettler:2010wi}.

Few mechanisms of realistic though very weak entropy production could take place during the
freeze-out of dark matter (DM) particles. But usually, the fraction of DM density was
quite low at the freezing our temperature and the effect is tiny.

An interesting effect, not covered in this paper, is the formation of bubbles walls that can take place in the early universe. The collision of them can lead to the formation of primordial black holes due to first order phase transition with background gravitational waves ~\cite{SR}, \cite{JN}

Most probably, the largest entropy production took place considering the SM during the EWPT. The entropy release happened when the universe went from a phase of symmetric electroweak phase to an asymmetric electroweak phase during the universe cooling. In the minimal SM with one Higgs field, the process is a mild crossover and the entropy production is about $13\%$~\cite{Chaudhuri:2017icn}.

According to the electroweak (EW) theory at the temperatures higher than a critical
one, $T > T_c$, the expectation value of the Higgs field, $\expval{\phi}$, in the fluid (plasma) is zero and the
universe is in electroweak symmetric phase~\cite{Bochkarev:1990fx}.
When the temperature drops below $\expval{T_c}$, a non-zero expectation value is
created, which gradually rises, with decreasing temperature, up to the vacuum expectation
value $\eta$. Such a state does not satisfy the conditions necessary for the
entropy conservation and an entropy production is expected.

A huge amount of entropy is released if EWPT is first order, which is the case even with the minimalist extension of standard model namely two-Higgs-doublet Model (2HDM). In what follows, we have considered a real 2HDM and scanned over certain parameter spaces and used numerical analysis to calculate the entropy production for some interesting and unique benchmark points. 

The paper is arranged as follows: In the next section details about 2HDM is given along with some LHC constrains
followed by the theoretical framework of the process. Due to cumbersome and very difficult analytical calculations, we did numerical analysis of certain parameters using BSMPT package~\cite{Basler:2020nrq} and it followed by a general discussion and conclusion. The paper has 2 Appendixes, giving details about the metric that is being used here and also the masses of the scalar bosons generated by 2HDM.

\section{2HDM: A small review}
There are two scalar doublets in the framework and they are defined as: 
\begin{equation} 
\varphi_{I} =  \left(
\begin{array}{c}
\phi_{I}^{+}\\
\frac{1}{\sqrt{2}}(v_{I} + \rho_{I}) + i\, \eta_{I}\\
\end{array}
\right),
\end{equation}
with $I=1,2$. Here $\phi^{\pm}_I$, $\rho_I$, $\eta_I$, and $v_I$ indicate the charged, neutral CP-even and neutral CP-odd degrees of freedom~(d.o.f.) and the vacuum expectation value~(vev) of the $I$-th doublet respectively.  

Prior to spontaneous symmetry breaking~(SSB), the tree-level 2HDM Lagrangian, assumes the form 
\bea
\label{lagrangian}
&&\mathcal{L} = \mathcal{L}_{\rm kin} + \mathcal{L}_{\rm Yuk} - V(\varphi_{1},\varphi_{2}) + \mathcal{L}_{6},
\eea
where, 
\bea
\label{lagrangianterms}
\mathcal{L}_{\rm kin} &=&  -\frac{1}{4} \sum_{X = {G^{a}},W^{i},B} X_{\m\n} X^{\m\n} + \sum_{I = 1,2} |D_{\m} \varphi_{I}|^2 + \sum_{\psi = Q,L,u,d,l} \bar{\psi} i \slashed{D} \psi,\nn\\
\mathcal{L}_{\rm Yuk} &=& \sum_{I=1,2} Y^{e}_I \, \bar{l} \,e \varphi_{I} + \sum_{I=1,2} Y^{d}_I \, \bar{q} \, d \varphi_{I} + \sum_{I=1,2} Y^{u}_I \, \bar{q} \, u \tilde{\varphi}_{I}, \nn\\
V(\varphi_{1},\varphi_{2}) &=& m_{11}^2 |\varphi_{1}|^2
 + m_{22}^2 |\varphi_{2}|^2 - ( \m^2 \varphi_{1}^{\dagger} \varphi_{2} +  h.c.) + \lambda_1 |\varphi_{1}|^4 + \lambda_2 |\varphi_{2}|^4 + \lambda_{3} |\varphi_{1}|^2 |\varphi_{2}|^2 \nn \\
  &&+  \lambda_4 |\varphi_{1}^{\dagger} \varphi_{2}|^2 + \Big[ \Big( \frac{\lambda_5}{2} \varphi_{1}^{\dagger} \varphi_{2} + \lambda_6 |\varphi_{1}|^2 + \lambda_7 |\varphi_{2}|^2 \Big) \varphi_{1}^{\dagger} \varphi_{2}  + h.c.\Big], \nn\\.
  \eea

In this paper, we assume the CP-conserving 2HDM scenario, and hence $\lambda_{6,7} = 0$. 
The electroweak symmetry is broken by the vacuum expectation values~(vev), namely $v_1$ and $v_2$  corresponding to the two doublets $\varphi_{1,2}$ respectively. This leads to the mixing of same types of degrees of freedom of $\varphi_{1,2}$. 

After sponteneous symmetry breaking, the Yukawa sector of the 2HDM can be written as,
\bea
- \mathcal{L}_{\rm Yuk}& = &\frac{1}{\sqrt{2}} (\kappa_D s_{\beta-\alpha} + \rho_D c_{\beta-\alpha}) \bar{D} D h + \frac{1}{\sqrt{2}} (\kappa_D c_{\beta-\alpha} - \rho_D s_{\beta-\alpha}) \bar{D} D H \nn\\
&&+ \frac{1}{\sqrt{2}} (\kappa_U s_{\beta-\alpha} + \rho_U c_{\beta-\alpha}) \bar{U} U h + \frac{1}{\sqrt{2}} (\kappa_U c_{\beta-\alpha} - \rho_U s_{\beta-\alpha}) \bar{U} U H \nn\\
&&+ \frac{1}{\sqrt{2}} (\kappa_L s_{\beta-\alpha} - \rho_L c_{\beta-\alpha}) \bar{L} L h + \frac{1}{\sqrt{2}} (\kappa_L c_{\beta-\alpha} - \rho_L s_{\beta-\alpha}) \bar{L} L H \nn\\
&&- \frac{i}{\sqrt{2}} \bar{U} \g_5 \rho_U U A + \frac{i}{\sqrt{2}} \bar{D} \g_5 \rho_D D A + \frac{i}{\sqrt{2}} \bar{L} \g_5 \rho_L L A \nn\\
&&+\Big(\bar{U}(V_{\text{CKM}}\, \rho_D P_R - \rho_U V_{\text{CKM}} P_L) D H^{+} + \bar{\n} \rho_L P_R L H^{+} + \text{h.c.} \Big),
\label{coupmult}
\eea

The generation indices of the fermionic fields have been suppressed in eq.~\eqref{coupmult}.
The limit $\cos (\beta-\alpha) \rightarrow 0$ with heavy scalars can lead back to the standard model scenario.

For type-I 2HDM, where range is allowed to be $|\cos (\beta-\alpha)| \lesssim 0.4$.
 Among the tree-level couplings, the decays of new scalars, $AZh$ and $H^{\pm}hW^{\mp}$ are proportional to $\cos (\beta-\alpha)$, whereas $AZH$ and $H^{\pm}HW^{\mp}$ are proportional to $\sin (\beta-\alpha)$. 
It is possible to realize an exact alignment in  the multi-Higgs-doublet models in the framework of certain additional symmetries of the 2HDM potential~\cite{Dev:2014yca,Pilaftsis:2016erj,Das:2017zrm,Benakli:2018vjk,Pramanick:2017wry}. 
The impact of the $\cos (\beta-\alpha) - \tan \beta   $ plane has been discussed in ref.~\cite{Craig:2013hca}.
A hierarchical spectrum like $m_A > m_H \sim m_{H^{\pm}} \sim v$ can lead to a first order EWPT providing an explanation for the matter-antimatter asymmetry.

\section{EWPT theory in 2HDM }
The lagrangian density of the Electroweak theory (discussed in details in the previous section)
in 2HDM can be expressed as~\ct{Langacker:2009my}
\be
\mathcal{L}= \mathcal{L}_f  + \mathcal{L}_{\rm Yuk} + \mathcal{L}_{\rm gauge, kin}  + \mathcal{L}_{\rm Higgs} \label{Eq: Total lagrangian}
\ee
The first term on the right hand side, $ \mathcal{L}_f$, is the kinetic term for the fermion-fields
\be
 \mathcal{L}_f &=&\sum_{j}i \lt(\bar{\Psi}^{(j)}_L \slashed{D}\Psi^{(j)}_L + \bar{\Psi}^{(j)}_R \slashed{D}\Psi^{(j)}_R  \rt) \\
 &=&i\bar{\Psi}_L \gamma^\mu (\partial_\mu + ig W_\mu + i g'Y_L B_\mu) \Psi_L   \nonumber \\
&& + i\bar{\Psi}_R  \gamma^\mu (\partial_\mu + ig W_\mu + i g'Y_R B_\mu) \Psi_R     \label{Eq: Fermionic Lagrangian}
\ee
where subscript $L$ and $R$ represents the left and right chiral field of that Fermion and $\slashed{D}$ is the covariant derivative~\ct{Langacker:2009my} and $j$ runs over all fermionic species listed in Table.[\ref{Table: mass of all fermion}].


The second term of Eq.~\ref{Eq: Total lagrangian}, Yukawa interaction term (for details, see previous section), $\mathcal{L}_{\rm Yuk}$ is~\cite{Logan:2014jla}  
\be
\mathcal{L}_{\rm Yuk}= -\left[ y_e \bar{e_R} \Phi_a^\dagger L_L + y_e^* \bar{L_L} \Phi_a^\dagger e_R  + \cdots \right]
\ee
where $y_e$ is a complex dimensionless constant, $\Phi_a$ ($a=1,2$) is a $SU(2)_L$ doublet and for the Lagrangian to be gauge invariant it is coupled with another $SU(2)_L$ fermion $L_L$. $e_R$ is the right chiral electron field and the same goes for other fermions like quarks, neutrinos, etc.

The third term $\mathcal{L}_{\rm gauge, kin} $ represents $U(1)$ invariant  kinetic term of four gauge bosons ($W^i, \, i=1,2,3$ and $B$). It can be written as 
\begin{eqnarray}
\mathcal{L}_{\rm gauge, kin} = -\frac{1}{4}G^i_{\mu \nu}{G^i}^{\mu \nu}-\frac{1}{4}F^B_{\mu \nu}{F^B}^{\mu \nu}          \label{Gauge-kinetic Lagrangian}
\end{eqnarray}
where $G^i_{\mu \nu}=\partial_\mu W^i_\nu-\partial_\nu W^i_\mu - g \epsilon^{ijk}W_\mu^j W_\nu^k$ and $F^B_{\mu \nu}=\partial_\mu B_\nu-\partial_\nu B_\mu$. 

The lagragian desity for the doublet Higgs bosons is given by 
\begin{eqnarray}
\mathcal{L}_{\rm Higgs} &=& (D^\mu \Phi_1)^\dagger (D_\mu \Phi_2) + (D^\mu \Phi_1)^\dagger (D_\mu \Phi_2) -V_{\rm tot}(\Phi_1,\Phi_2)  \nonumber \\  
&=& \{(\partial_\mu  + ig T^i W^i_\mu + i g'Y B_\mu ) \Phi_1  \}^\dagger\{(\partial_\mu +  ig T^i W^i_\mu + i g'Y B_\mu ) \Phi_1  \}   \nonumber \\
&&+ \{(\partial_\mu  + ig T^i W^i_\mu + i g'Y B_\mu ) \Phi_2 \}^\dagger\{(\partial_\mu +  ig T^i W^i_\mu + i g'Y B_\mu ) \Phi_2 \}    - V_{\rm tot}(\Phi_1,\Phi_2, T)   \nonumber \\  \label{Eq:eq1}
\end{eqnarray} 

We define  
\begin{eqnarray}
	\mathcal{W}_\mu=gT^iW^i_\mu + g'Y B_\mu  \label{Eq: definition of W}
\end{eqnarray}
Thus from Eq.\ref{Eq:eq1}, we get
\begin{eqnarray}
	 \mathcal{L}_{\rm Higgs,kin}=(\partial^\mu {\Phi_a}^\dagger )(\partial_\mu \Phi_a) - i ({\mathcal{W}^\mu}\Phi_a)^\dagger (\partial_\mu \Phi_a) + i (\partial^\mu{\Phi_a}^\dagger)\mathcal{W}_\mu\Phi_a+  (\mathcal{W}^\mu\Phi_a)^\dagger \mathcal{W}_\mu\Phi_a     
\end{eqnarray}

The standard CP-conserving 2HDM potential $V_{\rm tot}(\Phi_1,\Phi_2, T)$ consists of tree level potential $V_{\rm tree}(\Phi_1,\Phi_2)$ 
\begin{eqnarray} 
V_{\rm tree}(\Phi_1,\Phi_2)   =&& m_{11}^2 \Phi_1^\dagger \Phi_1 + m_{22}^2 \Phi_2^\dagger \Phi_2 - \left[m_{12}^2 \Phi_1^\dagger \Phi_2 + m_{12}^* \Phi_2^\dagger \Phi_1 \right] + \frac{1}{2} \lambda_1 \left(\Phi_1^\dagger \Phi_1\right)^2 \nonumber \\ 
&&+  \frac{1}{2} \lambda_2 \left(\Phi_2^\dagger \Phi_2\right)^2  + \lambda_3 \left(\Phi_1^\dagger \Phi_1\right)\left(\Phi_2^\dagger \Phi_2\right) + \lambda_4 \left(\Phi_1^\dagger \Phi_2\right)\left(\Phi_2^\dagger \Phi_1\right) \nonumber \\
&&+ \left[\frac{1}{2} \lambda_5  \left(\Phi_1^\dagger \Phi_2\right)^2 + \frac{1}{2} \lambda_5^*  \left(\Phi_2^\dagger \Phi_1\right)^2 \right]
\end{eqnarray}
and other correction terms $V_{\rm CW}{\left(v_1,v_2 \right)}$ and $V_T$. The correction terms are defined as~\cite{Blinov:2015vma, Basler:2016obg}


\be
V_{\rm CW}\left(v_1+v_2 \right)=\sum_i \frac{n_i}{64\pi^2} (-1)^{2s_i}m_i^4\left(v_1,v_2\right)\left[ \log\left( \frac{m_i^2 \left(v_1,v_2 \right)}{\mu^2} \right) - c_i \right] \\
V_T=\frac{T^4}{2\pi^2}\left( \sum_{i={\rm bosons}} n_i J_{B}\left[\frac{m_i^2(v_1,v_2)}{T^2}\right]  + \sum_{i={\rm fermions}} n_i J_{F}\left[\frac{m_i^2(v_1,v_2)}{T^2}    \right] \right)
\ee
where $\mu$ is the  renormalisation scale which we take to be $246~GeV$.

The potential dependent mass of fermions and bosons $m_i\left(v_1+v_2 \right)$ and the corresponding $n_i$, $s_i$, and $c_i$ are discussed  in details in Appendix~\ref{s-a}.

$J_B$ and $J_F$ are approximated Landau gauge up to leading orders as following  

\be
T^4 J_B \left[\frac{m^2}{T} \right] &&= -\frac{\pi^4 T^4}{45} + \frac{\pi^2}{12}T^2 m^2 - \frac{\pi}{6}T (m^2)^{3/2} - \frac{1}{32}m^4 \ln \frac{m^2}{a_b T^2} + \cdots   \\
T^4 J_F \left[\frac{m^2}{T} \right] &&=  \frac{7\pi^4 T^4}{360} - \frac{\pi^2}{24}T^2 m^2 - \frac{1}{32}m^4 \ln \frac{m^2}{a_f T^2} + \cdots 
\ee
where $a_b=16a_f=16\pi^2 \exp(3/2-2\gamma_E)$ with $\gamma_E$ being the Euler-Mascheroni constant.


When the temperature of the universe drops down to the critical temperature $T_c$, a second local minimum appears with the same height of the global minimum situated at
 $\expval{\Phi_1}=\expval{\Phi_2}=0$~\cite{Katz:2014bha}. The critical temperature can be obtained using the following expression: 
\be
V_{\rm tot}\left(\Phi_1=0,\Phi_2=0,T_c \right)= V_{\rm tot}\left(\Phi_1=v_1,\Phi_2=v_2,T_c \right)
\ee

During Electroweak Phase Transition (EWPT), if $\rho$ and $P$ are respectively the energy density and pressure of the 
fluid determining the course of evolution of the early universe , then from~[\ref{Eq: Total lagrangian}] 

\be
\rho &=& \rho_{f} + \rho_{\rm gauge, kin} + \rho_{Higgs} - g^{00}\mathcal{L}_{\rm Yuk}  \\
P &=& P_{f} + P_{\rm gauge, kin} + P_{Higgs} - \frac{1}{3}g^{ii}\mathcal{L}_{\rm Yuk}
\ee
We have assumed that dark matter and other components might have been present but they did not contribute much to the energy density of the universe during the particular epoch of EWPT which happened in radiation domination. The expressions for $\rho_{f},  P_{f}$ and $\rho_{\rm gauge, kin}, P_{\rm gauge, kin}$ appear solely from fermionic and gauge sectors and their interactions. The Stress-energy tensor for the above quantities is mentioned in the Appendix~\ref{Appendix: Energy-momentum tensor}. 

\be 
\rho_\text{\tiny H+F+G} &&=\left[ \{  \partial^0 \Phi_a^\dagger -  i ({\mathcal{W}^0}\Phi_a)^\dagger \}\partial^0 \Phi_a + \{\partial^0\Phi_a + i \mathcal{W}^0\Phi_a     \}\partial^0 \Phi_a^\dagger \right]  \nonumber \\
&& -g^{00} \Big[(\partial^\alpha {\Phi_a}^\dagger )(\partial_\alpha\Phi_a) - i ({\mathcal{W}^\alpha}\Phi_a)^\dagger (\partial_\alpha \Phi_a) + i (\partial^\alpha{\Phi_a}^\dagger)\mathcal{W}_\alpha\Phi_a+  (\mathcal{W}^\alpha\Phi_a)^\dagger \mathcal{W}_\alpha\Phi_a \Big]    \nonumber \\
&&  -g^{00} \left[  V_{\rm tot}(\Phi_1,\Phi_2, T)    \right] 
\ee

\be
P_\text{\tiny H+F+G} &&=\left[ \{  \partial^q \Phi_a^\dagger -  i ({\mathcal{W}^q}\Phi_a)^\dagger \}\partial^q \Phi_a + \{\partial^q\Phi_a + i \mathcal{W}^q\Phi_a     \}\partial^q \Phi_a^\dagger \right]  \nonumber \\
&& -g^{q q} \Big[(\partial^\alpha {\Phi_a}^\dagger )(\partial_\alpha\Phi_a) - i ({\mathcal{W}^\alpha}\Phi_a)^\dagger (\partial_\alpha \Phi_a) + i (\partial^\alpha{\Phi_a}^\dagger)\mathcal{W}_\alpha\Phi_a+  (\mathcal{W}^\alpha\Phi_a)^\dagger \mathcal{W}_\alpha\Phi_a \Big]    \nonumber \\
&&  -g^{q q} \left[  V_{\rm tot}(\Phi_1,\Phi_2, T)\right]
\label{Faltu equation 2}
\ee

The early universe was flat, hence the metric $g_{\mu \nu}=(+,-,-,-)$. And hence $\rho_\text{\tiny H+F+G} + P_\text{\tiny H+F+G}$ becomes:
\be
\rho_\text{\tiny H+F+G} + P_\text{\tiny H+F+G}&&= \partial^0\Phi_a \partial^0 \Phi_a^\dagger + 2\left[  (\mathcal{W}^\alpha\Phi_a)^\dagger \mathcal{W}_\alpha\Phi_a  \right]\nonumber \\
&& + \Big[(\partial^0 {\Phi_a}^\dagger )(\partial_0\Phi_a) - i ({\mathcal{W}^0}\Phi_a)^\dagger (\partial_0 \Phi_a) + i (\partial^0{\Phi_a}^\dagger)\mathcal{W}_0\Phi_a \Big]    \nonumber \\
\label{Faltu equation 5}
\ee
where the explicit expression for $\rho_\text{\tiny H+F+G}$ is given in the Eq.\ref{Faltu equation 4}
\be
\rho_\text{\tiny H+F+G} &&=\partial^0\Phi_a \partial^0 \Phi_a^\dagger + \left[   (\mathcal{W}^\alpha\Phi_a)^\dagger \mathcal{W}_\alpha\Phi_a  \right] \nonumber \\
&&- \left[  V_{\rm tot}(\Phi_1,\Phi_2, T)   +   \mathcal{L}_{\rm Yuk}  \right]  \label{Faltu equation 4}
\ee

The oscillations of  the Higgs fields around minimum after it appeared in the course of the phase transition are damped due to particle 
production by the oscillating field. The characteristic time is equal to the decay width of the Higgses and it is large in comparison with
the expansion rate and the universe cooling rate. So we may assume that Higgses essentially live in the minimum of the potential. In
principle, it can be calculated numerically by the solution of the corresponding Klein-Gordon equation with damping induced by the
particle production.~\cite{Chaudhuri:2017icn}.

With the above assumption
\be \label{22}
\rho &=& \Dot{\Phi}_{a, {\rm min}}^2
+  V_{\rm tot}(\Phi_1,\Phi_2, T) + \frac{g_* \pi^2}{30} T^4
\ee

The last term in Eq.\ref{22} arises from the Yukawa interaction between fermions and Higgs bosons and from
the energy density of the fermions, the Gauge bosons and the interaction between the Higgs and Gauge bosons. This is the energy density of the relativistic particles which have not gained mass till the moment of EWPT.

Since for relativistic species $P=(1/3) \rho$, we can write
\be
P= \frac{\Dot{\Phi}_{a, {\rm min}}^2}{2} 
+ \frac{1}{3}\frac{g_* \pi^2}{30} T^4
\ee

The oscillations of scalar fields around their minimas are quickly damped, so we take the time derivative of the fields equivalent to the their derivative around the minimas, and neglect higher order terms of their time derivative $\Dot{\phi}^2$ and so on. And as a result the evolution of the minimas induced by the expansion of the universe is very slow.

The entropy conservation law holds when the plasma (assumed to be an ideal fluid) was in thermal equilibrium with negligible chemical potential. But as the temperature went below $T_C$, EWPT happened and the universe went into a thermally non-equilibrium state. It is to be noted that one of the main consequences of EWPT is Electroweak baryogenesis and following Sakharov's principle, out of equilibrium process is a necessary condition for successful baryogenesis.

As a result of this deviation from thermal equilibrium, the entropy conservation law is no more valid during EWPT and hence a rise in the entropy production can be noticed significantly during this process. 

To calculate this production, it is necessary to solve the evolution equation for energy density conservation
\be \label{fried}
\Dot{\rho}=-3H(\rho+P)
\ee

From hence forward computational analysis was used for further calculations which are discussed in the next section.

\section{Entropy release in 2HDM scenarios}
At very early time when the temperature of universe $T\gg T_c$, the universe was in thermal equilibrium and also was dominated by relativistic species. Almost all of the fermions and bosons were massless and contribution from those who were already massive (e.g. DM) to the total energy density of the universe was insignificant. During that epoch the chemical potential of the massless bosons was zero and with the assumption that chemical potential of the fermions was negligible, 
the entropy density per comoving volume was conserved and is given by

\begin{eqnarray}
\label{eq:entropy-conservation}
s\equiv\frac{\rho_r +P_r}{T}a^3 = {\rm const.}
\end{eqnarray}
where the subscript $r$ is used to indicate relativistic components. For our scenario
\begin{eqnarray}
\label{eq:rho+P}
\rho_r +P_r \sim g_\star T^4
\end{eqnarray}
$g_\star$ is not constant over time; it depends on the components of the hot primordial hot soup. Those two equations (Eq.~\eqref{eq:entropy-conservation} and Eq.~\eqref{eq:rho+P}) implies
\begin{eqnarray}
T \sim a^{-1}
\end{eqnarray}

As long as the thermal equilibrium were maintained, $s$ remained constant. If the thermal equilibrium was not remained at some epoch at later time, the value of $s$ and thus $g_\star \left(T\right) a^3 T^3$, might have increased as entropy can only either increase or remain constant.

As temperature decreased to $T_c$ Higgs potential got degenerate minima. Later temperature dropped down more. If the temperature dropped to the mass of any component of the relativistic plasma, that component gained mass and became non-relativistic and decoupled from relativistic fluid.  We are assuming this process was instantaneous  and 
 the universe was not in thermal equilibrium.There is change in $g_\star$ of the relativistic plasma. This led to increase of $s$. If this decoupling process was in thermal equilibrium, it would cause a sudden increase of temperature of the universe.
 
Suppose at $T_c$, the scale factor was $a_c$ and $g_\star\equiv g_{\star,{\rm tot}}=110.75$ for our 2HDM model and thus at that moment $s_c\sim g_{\star,{\rm tot}} \, a_c \, T_c$. While temperature dropped to $T\sim T_x$, the component '$x$' would decoupled and thus $g_\star$ of the relativistic plasma would decrease. If the instantaneous decoupling process was occurred at equilibrium, it would increase the temperature of the photons However, Why are we considering as we are considering non-equilibrium case, $s$ of the universe was increased. If $g_{\star,+}$ and $g_{\star,-}$ be the $g_\star$-factor before and after the decoupling of the '$x$', then change in $s$ relative to the time of critical temperature
\begin{eqnarray}
\label{eq:definition of entropy change}
\frac{\delta s}{s_c}= \frac{\left(g_{\star,+} \, a_x \, T_x\right)^3 - \left(g_{\star,{\rm tot}} \, a_c \, T_c\right)^3}{\left(  g_{\star,{\rm tot}} \, a_c \, T_c \right)^3}
\end{eqnarray}

BSMPT is a C++ package which deals with various properties and features related to 2HDM and baryon asymmetry. In this case, the package was used to calculate the critical temperature for $T_c$ and the vacuum expectation value vev and also $V_{\rm eff}(T)$ for each benchmark points. The calculation was repeated for $4$ parameter space, the first one being the benchmark points provided in the BSMPT manual. The differential equation Eq.\ref{fried} was solved numerically by interpolating the data for $V_{\rm eff}(T)$ for all the benchmark points and the entropy release was calculated for the same.

\begin{figure}[h!]
\centering
\includegraphics[]{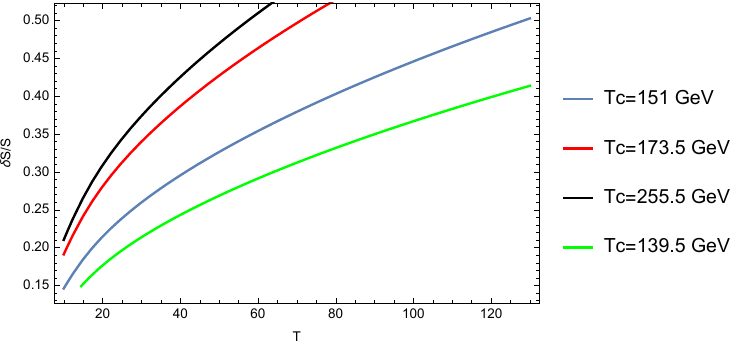}
\caption{Entropy production for various benchmark points as given in Table-\ref{Table:1 Benchmark values}.  }
\label{f-entropy-1}
\end{figure}

For 4 different benchmark points, as mentioned in Table\ref{Table:1 Benchmark values} the entropy release has been calculated with the assumption of $a_cT_c \sim 1$ and it is presented in Fig. \ref{f-entropy-1}

\begin{table}[h!]
    \centering
\begin{center}
\caption{2HDM Benchmark points for entropy production}\label{Table:1 Benchmark values}
\begin{tabular}{ |c|c|c|c|c|} 
\hline
& $m_h$& $m_H$ & $m_{H^{\pm}}=m_A$ & $\tan\beta$ \\
\hline
Benchmark-I & $125$ & $500$ & $500$ &  $10$\\
\hline
Benchmark-II & $125$ & $400$ & $500$ & $10$\\
\hline
Benchmark-III & $125$ & $90$ & $400$ &  $10$\\
\hline
Benchmark-IV & $125.09$ & $228.17 $ & $233$ &  $6.94$\\
\hline
\end{tabular}
\end{center}
%
\vspace{-0.8cm}
\begin{center}
\begin{tabular}{|c|c|c|c|c|c|c|}
\hline
 &$\lambda_1$ & $\lambda_2$ & $\lambda_3$& $\lambda_4$ & $\lambda_5$ &  $T_c$ \\
\hline
Benchmark-I &$4.13$ & $0.29$ & $4.15$ & $0$ & $0$ &  $T_c=255.5~GeV$\\
\hline
Benchmark-II & $0.25$ & $0.25$ & $12.65$ & $-1.48$ & $-1.48$& $T_c=173.5~GeV$ \\
\hline
Benchmark-III & $0.133$ & $0.259$ & $5.02$ & $-2.51$ & $-2.51$& $T_c=151~GeV$\\
\hline
Benchmark-IV & $1.22$ & $0.29$ & $-0.51$ & $4.07$ & $-3.86$ & $T_c=139.5~GeV$ \\
\hline
\end{tabular}
\end{center}
\end{table}

As it is seen from Fig.\ref{f-entropy-1}, the amount of entropy release increases as the critical temperature for EWPT increases. For example, the entropy production for $T_c=139.5 GeV$ is $\sim 41 \%$, for $T_c=151 GeV$ is $\sim 52 \%$, for $T_c=173.5 GeV$ is $\sim 63 \%$ and for $T_c=255.5 GeV$ is $\sim 73 \%$. All these results are way higher than the entropy release by EWPT in SM which is $\sim 13\%$~\cite{Chaudhuri:2020wjo}.

The main reason for this excess in the production of entropy is the extra scalar bosons produced in 2HDM which contributes the most to the process. It is to be noted that the contributions from lighter particles like electrons and neutrinos are similar to that of SM.

\section{Conclusion}
It is shown and calculated in this paper that the total entropy release due to EWPT is very large even in the framework of minimal extension of SM of particle physics namely 2HDM compared to minimal SM of physics. It is a proven fact that unlike SM where EWPT is of second order, in the mere extension of SM EWPT becomes a first order phase transition. An interesting fact is that as $g_*$ decreases as the temperature falls down. But as we go to a very low temperature scale, the minimum temperature ($T_{min}$) takes the value of the particle mass and their contribution remains the same like that of SM. 

There are two points which should be noted. Firstly, the benchmark points are unique and they were calculated using the BSMPT package. BSMPT gives the results for those benchmark points which satisfies the condition vev$/T_c >1$. All the benchmark points used here satisfy this condition. Secondly in this paper we have considered only the real sector of 2HDM. If other extensions of 2HDM like complex 2HDM are considered, there might be considerable change in the entropy production. 

In passing by, two effects are needed to be mentioned even though they are beyond the scope of this paper. Firstly, the entropy release due to EWPT can considerably reduce the abundance of Dark matter present in the universe before EWPT. Detailed calculations of this dilution factor for Standard model are done in ~\cite{Chaudhuri:2020wjo}. Secondly, bubble walls that were formed might collide and may produce primordial black holes and might lead to a sufficient entropy production. The bubble collisions are also the source of primordial gravitational wave background. These will be studied in the subsequent papers. 

\vspace{20pt}
\noindent{\bf Author Contribution}

\noindent
Article by A.C. and M.K.. The authors contributed equally to this~work. All authors have read and agreed to the published version of the manuscript.

\vspace{20pt}

\noindent {\bf Acknowledgement}

\noindent 
The work of AC is funded by RSF Grant 19-42-02004. The work by M.K. has been performed with a support of the Ministry of Science and Higher Education of the Russian Federation, Project "Fundamental problems of cosmic rays and dark matter", No 0723-2020-0040.


{}

\newpage


\appendix
\section{Appendix: Energy Momentum Tensor  \label{Appendix: Energy-momentum tensor}}
\be
T^{\mu \nu}_f &=& \sum_{j}i \lt(\bar{\Psi}^{(j)}_L \gamma^{\mu} \partial^\nu \bar{\Psi}^{(j)}_L + \bar{\Psi}^{(j)}_R  \gamma^\mu \partial^\nu \bar{\Psi}^{(j)}_R\rt)  - g^{\mu \nu}\mathcal{L}_f       \\
T^{\mu \nu}_{\rm gauge, kin}&=&  +  \lt[F^{B \, \mu \alpha} \partial^\nu B_\alpha -  \frac{1}{4} g^{\mu \nu} F^B_{\alpha \beta} F^{B \, \alpha \beta}\rt]  \nonumber \\
&&+   \lt[G^{i \, \mu \alpha} \partial^\nu W_\alpha -  \frac{1}{4} g^{\mu \nu} G^i_{\alpha \beta} G^{i \,\alpha \beta}\rt]  - g \epsilon^{ijk} \left( W^{\mu j} W^k_{\alpha} \partial^\nu W^\alpha - W^j_\alpha W^{\nu k} \partial^\mu W^\alpha \right)    \nonumber \\
\ee

\section{Appendix: Masses of new Scalars  \label{s-a}}
\be
c_i=
\begin{cases}
      \frac{5}{6}, & \left(i=W^\pm, Z, \gamma \right)\\
      \frac{3}{2}, & \text{otherwise}
  \end{cases}
\ee

\begin{center}
\begin{tabular}{ | c | c| c | c| c| } 
\hline
Bosons & $n_i$ & $s_i$ & $m(v)^2$ & \\
\hline
$h$ & $1$ & $1$ & eigenvalues of \ref{Mass_of_neutral_Higgs_bosons} &  Higgs \\
\hline
$H$ & $1$ & $1$ & eigenvalues of \ref{Mass_of_neutral_Higgs_bosons} & Higgs \\
\hline
$A$& $1$ & $1$ & eigenvalues of \ref{Mass_of_neutral_Higgs_bosons} & Higgs\\
\hline
$G^0$& $1$ & $1$ & eigenvalues of \ref{Mass_of_neutral_Higgs_bosons} & Goldstone\\
\hline
$H^\pm$& $2$ & $1$ & Eq.\ref{Mass_of_mHPlusMinus}& Charged  Higgs\\
\hline
$G^\pm$& $2$ & $1$ & Eq.\ref{Mass_of_mGPlusMinus}& Charged Goldstone\\
\hline
$Z_L$& $1$ & $1$ & Eq.\ref{Mass_of_Z} & Higgs\\
\hline
$Z_T$& $2$ & $2$ & Eq.\ref{Mass_of_Z} & Higgs\\
\hline
$W_L$& $2$ & $1$ & Eq.\ref{Mass_of_W} & Higgs\\
\hline
$W_T$& $4$ & $2$ & Eq.\ref{Mass_of_W} & Higgs\\
\hline
$\gamma_L$& $1$ & $2$ & Eq.\ref{Mass_of_gamma} & \\
\hline
$\gamma_T$& $2$ & $2$ & Eq.\ref{Mass_of_gamma} & \\
\hline
\end{tabular}
\end{center}

\be
&& m_W^2=\frac{g^2}{4}v^2. \label{Mass_of_W}\\
&& m_Z^2=\frac{g^2+g'^2}{4}v^2. \label{Mass_of_Z} \\
&& m_\gamma^2=0. \label{Mass_of_gamma}
\ee

\be
\Bar{m}_{H^\pm}^2 &&=\frac{1}{2} \left( \mathcal{M}_{11}^C + \mathcal{M}_{22}^C \right)  +\frac{1}{2} \sqrt{4\left( \left( \mathcal{M}_{12}^C\right)^2 + \left( \mathcal{M}_{13}^C \right)^2 \right)+\left(\mathcal{M}_{11}^C-\mathcal{M}_{22}^C \right)^2}.  \label{Mass_of_mHPlusMinus} \\
\Bar{m}_{G^\pm}^2 &&=\frac{1}{2}\left( \mathcal{M}_{11}^C + \mathcal{M}_{22}^C +  \right) -\frac{1}{2} \sqrt{4\left( \left( \mathcal{M}_{12}^C\right)^2 + \left( \mathcal{M}_{13}^C \right)^2\right) +\left(\mathcal{M}_{11}^C-\mathcal{M}_{22}^C  \right)^2 }.    \label{Mass_of_mGPlusMinus}
\ee
where
\be
c_1 &&=\frac{1}{48}\left(12\lambda_1 + 8 \lambda_3 + 4 \lambda_4 + 3 \left( 3g^2 + g'^2\right)\right) \\
c_2 &&= \frac{1}{48}\left(12\lambda_2 + 8 \lambda_3 + 4 \lambda_4 + 3 \left( 3g^2 + g'^2\right) + \frac{24}{v_2^2}m_t^2(T=0)\right) \nonumber \\
&&+\frac{1}{2v_2^2}m_b^2(T=0)
\ee
where $m_t(T=0)=172.5 {\rm Gev}$ and $m_{b}(T=0)=4.92 {\rm GeV}$.
For our case $(v_3=0)$,
\be
\mathcal{M}_{11}^C &&=m_{11}^2 +\lambda_1 \frac{v_1^2}{2} + \lambda_3 \frac{v_2^2}{2}  \\
\mathcal{M}_{22}^C &&=m_{22}^2 +\lambda_2 \frac{v_2^2}{2} + \lambda_3 \frac{v_1^2}{2}  \\
\mathcal{M}_{12}^C &&=\frac{v_1v_2}{2}\left(\lambda_4+\lambda_5\right)-m_{12}^2 \\
\mathcal{M}_{13}^C &&=0
\ee

Masses of $h$, $H$ and $A$ are the eigen values of the matrix
\be
\Bar{\mathcal{M}}^N=\left( \mathcal{M}^N \right) \label{Mass_of_neutral_Higgs_bosons} 
\ee
For our case $(v_3=0)$,
\be
\mathcal{M}_{11}^N &&= m_{11}^2+\frac{3\lambda_1}{2}v_1^2+ \frac{\lambda_3 +\lambda_4}{2} v_2^2 + \frac{1}{2} \lambda_5 v_2^2\\
\mathcal{M}_{22}^N &&= m_{11}^2+\frac{\lambda_1}{2}v_1^2+ \frac{\lambda_3 +\lambda_4}{2} v_2^2 - \frac{1}{2} \lambda_5 v_2^2\\
\mathcal{M}_{33}^N &&= m_{22}^2+\frac{3\lambda_2}{2}v_2^2 + \frac{1}{2}\left(\lambda_3+\lambda_4 +\lambda_5 \right)v_1^2\\
\mathcal{M}_{44}^N &&= m_{22}^2+\frac{\lambda_2}{2}v_2^2 + \frac{1}{2}\left(\lambda_3+\lambda_4 -\lambda_5 \right)v_1^2\\
\mathcal{M}_{12}^N &&= 0\\
\mathcal{M}_{13}^N &&=-m_{12}^2+\left(\lambda_3+\lambda_4+\lambda_5\right) v_1 v_2 \\
\mathcal{M}_{14}^N &&=0 \\
\mathcal{M}_{23}^N &&=0 \\
\mathcal{M}_{24}^N &&= -m_{12}^2+\lambda_5 v_1 v_2\\
\mathcal{M}_{34}^N &&= 0
\ee

\begin{table}
\caption{Field dependent mass of all fermions}\label{Table: mass of all fermion}
\begin{center}
\begin{tabular}{ | c | c| c | c| c| }  
\hline
Fermions & $n_i$ & $s_i$ & $m_f(T=0)$ & \\
\hline
$e$ & $4$ & $\frac{1}{2}$ & $\frac{y_e}{\sqrt{2}}v_k$ & lepton\\

$\mu$ & $4$  & $\frac{1}{2}$ & $\frac{y_\mu}{\sqrt{2}}v_k$ & lepton\\

$\tau$ & $4$ & $\frac{1}{2}$ & $\frac{y_\tau}{\sqrt{2}}v_k$ & lepton\\
\hline
$u$ & $12$ & $\frac{1}{2}$ & $\frac{y_u}{\sqrt{2}}v_k$ & quark\\
$c$ & $12$ & $\frac{1}{2}$ & $\frac{y_c}{\sqrt{2}}v_k$ & quark\\
$t$ & $12$  & $\frac{1}{2}$ & $\frac{y_t}{\sqrt{2}}v_k$ & quark\\
$d$ & $12$ & $\frac{1}{2}$ & $\frac{y_d}{\sqrt{2}}v_k$ & quark\\
$s$ & $12$ & $\frac{1}{2}$ & $\frac{y_s}{\sqrt{2}}v_k$ & quark\\
$b$ & $12$ & $\frac{1}{2}$ & $\frac{y_b}{\sqrt{2}}v_k$ & quark\\
\hline
\end{tabular}
\end{center}
\end{table}

\begin{thebibliography}{99}

\bibitem{Gorbunov:2011zzc}
D.~S.~Gorbunov and V.~A.~Rubakov,
``Introduction to the theory of the early universe: Cosmological perturbations and inflationary theory,''

\bibitem{Bambi:2015mba}
C.~Bambi and A.~D.~Dolgov,
``Introduction to Particle Cosmology,''

\bibitem{Dolgov:2000ht}
A.~D.~Dolgov, P.~D.~Naselsky and I.~D.~Novikov,
``Gravitational waves, baryogenesis, and dark matter from primordial black holes,''
[\href{https://arxiv.org/abs/astro-ph/0009407}{{\shilacred \ttfamily arXiv:astro-ph/0009407 [astro-ph]}}].



\bibitem{Chaudhuri:2020wjo}
A.~Chaudhuri and A.~Dolgov,
``PBH evaporation, baryon asymmetry,and dark matter,''
[\href{https://arxiv.org/abs/2001.11219}{{ \shilacred \ttfamily arXiv:2001.11219 [astro-ph.CO]}}].
\bibitem{khlopovPPNP} M.Khlopov,'' What comes after the Standard model?'' Progress in Particle and Nuclear Physics 116 (2021) 103824; https://doi.org/10.1016/j.ppnp.2020.103824
\bibitem{Schettler:2010wi}
T.~Boeckel, S.~Schettler and J.~Schaffner-Bielich,
``The Cosmological QCD Phase Transition Revisited,''
\href{https://doi.org/10.1016/j.ppnp.2011.01.017}{ \tcmag{Prog. Part. Nucl. Phys. \textbf{66}, 266-270 (2011)}}
[\href{https://arxiv.org/abs/1012.3342}{{\shilacred \ttfamily arXiv:1012.3342 [astro-ph.CO]}}].

\bibitem{Bochkarev:1990fx}
A.~I.~Bochkarev, S.~V.~Kuzmin and M.~E.~Shaposhnikov,
``Electroweak baryogenesis and the Higgs boson mass problem,''
\href{https://doi.org/10.1016/0370-2693(90)90069-I}{ \tcmag{Phys. Lett. B \textbf{244}, 275-278 (1990)}}

\bibitem{SR}
Konoplich R.V., Rubin S.G., Khlopov M.Yu. , "Formation of black holes in first-order phase transitions in the Universe", {\tcmag{ Astronomy Letters 24(4):413-417}}.

\bibitem{JN}
Jedamzik K. and Niemeyer J.C., "Primordial Black Hole Formation during First-Order Phase Transitions", Phys.Rev.D 59 (1999) 124014,  arXiv:astro-ph/9901293.


\bibitem{Basler:2018cwe}
P.~Basler and M.~M\"uhlleitner,
``BSMPT (Beyond the Standard Model Phase Transitions): A tool for the electroweak phase transition in extended Higgs sectors,''
\href{https://doi.org/10.1016/j.cpc.2018.11.006}{ \tcmag{Comput. Phys. Commun. \textbf{237}, 62-85 (2019)}}
[\href{https://arxiv.org/abs/1803.02846}{{\shilacred \ttfamily arXiv:1803.02846 [hep-ph]}}].

\bibitem{Basler:2020nrq}
P.~Basler, M.~Muhlleitner and J.~M\"uller,
``BSMPT v2 A Tool for the Electroweak Phase Transition and the Baryon Asymmetry of the Universe in Extended Higgs Sectors,''
[\href{https://arxiv.org/abs/2007.01725}{{\shilacred \ttfamily arXiv:2007.01725 [hep-ph]}}].


\bi{Langacker:2009my}
P.~Langacker,
``Introduction to the Standard Model and Electroweak Physics,''
\href{https://doi.org/10.1142/9789812838360_0001}{\tcmag{ In The Dawn Of The LHC Era: TASI 2008 (3-48).}}
[\href{https://arxiv.org/abs/0901.0241}{{ \shilacred \ttfamily arXiv:0901.0241 [hep-ph]}}].


\bibitem{Logan:2014jla}
H.~E.~Logan,
``TASI 2013 lectures on Higgs physics within and beyond the Standard Model,''
[\href{https://arxiv.org/abs/1406.1786}{{ \shilacred \ttfamily arXiv:1406.1786 [hep-ph]}}].


\bibitem{Blinov:2015vma}
N.~Blinov, S.~Profumo and T.~Stefaniak,
``The Electroweak Phase Transition in the Inert Doublet Model,''
 \href{https://doi.org/10.1088/1475-7516/2015/07/028}{\tcmag{  JCAP \textbf{07}, 028 (2015)}}
[\href{https://arxiv.org/abs/1504.05949}{{ \shilacred \ttfamily arXiv:1504.05949 [hep-ph]}}].

\bibitem{Basler:2016obg}
P.~Basler, M.~Krause, M.~Muhlleitner, J.~Wittbrodt and A.~Wlotzka,
``Strong First Order Electroweak Phase Transition in the CP-Conserving 2HDM Revisited,''
\href{https://doi.org/10.1007/JHEP02(2017)121}{ \tcmag{ 
JHEP \textbf{02}, 121 (2017)}}
[\href{https://arxiv.org/abs/1612.04086}{{ \shilacred \ttfamily arXiv:1612.04086 [hep-ph]}}].


\bibitem{Katz:2014bha}
A.~Katz and M.~Perelstein,
``Higgs Couplings and Electroweak Phase Transition,''
\href{https://doi.org/10.1007/JHEP07(2014)108}{\tcmag{ JHEP \textbf{07}, 108 (2014)}}
[\href{https://arxiv.org/abs/1401.1827}{{\shilacred \ttfamily arXiv:1401.1827 [hep-ph]}}].

\bibitem{Chaudhuri:2017icn} 
  A.~Chaudhuri and A.~Dolgov,
  ``Electroweak phase transition and entropy release in the early universe,''
  \href{https://doi.org/10.1088/1475-7516/2018/01/032}{\tcmag{  JCAP {\bf 1801}, 032 (2018)}}
  [\href{https://arxiv.org/abs/1711.01801}{{\shilacred \ttfamily arXiv:1711.01801 [hep-ph]}}].
  
\bibitem{Lahiri}
Lahiri, A., \& Pal, P. B. (2005). A first book of quantum field theory. CRC Press.


\bibitem{Karmakar:2019vnq}
S.~Karmakar and S.~Rakshit,
``Relaxed constraints on the heavy scalar masses in 2HDM,''
 \href{https://doi.org/10.1103/PhysRevD.100.055016}{\tcmag{ Phys. Rev. D \textbf{100}, no.5, 055016 (2019)}}
[\href{https://arxiv.org/abs/1901.11361}{{\shilacred \ttfamily arXiv:1901.11361 [hep-ph]}}].

\bibitem{Chakraborty:2015raa}
I.~Chakraborty and A.~Kundu,
``Scalar potential of two-Higgs doublet models,''
\href{https://doi.org/10.1103/PhysRevD.92.095023}{ \tcmag{Phys. Rev. D \textbf{92}, no.9, 095023 (2015)}}
[ arXiv:1508.00702 [hep-ph]].


\bibitem{Athron:2020sbe}
P.~Athron, C.~Bal\'azs, A.~Fowlie and Y.~Zhang,
``PhaseTracer: tracing cosmological phases and calculating transition properties,''
\href{https://doi.org/10.1140/epjc/s10052-020-8035-2}{ \tcmag{Eur. Phys. J. C \textbf{80}, no.6, 567 (2020)}}
[ arXiv:2003.02859 [hep-ph]].

\bibitem{Ginzburg:2004vp}
  I.~F.~Ginzburg and M.~Krawczyk,
  ``Symmetries of two Higgs doublet model and CP violation,''
  \href{https://doi.org/10.1103/PhysRevD.72.115013}{ \tcmag{Phys.\ Rev.\ D {\bf 72} (2005) 115013}}
  [ arXiv:hep-ph/0408011 [hep-ph]].

\bibitem{ElKaffas:2007rq}
  A.~W.~El Kaffas, P.~Osland and O.~M.~Ogreid,
  ``CP violation, stability and unitarity of the two Higgs doublet model,''
  Nonlin.\ Phenom.\ Complex Syst.\  {\bf 10} (2007) 347
  [\href{https://arxiv.org/abs/hep-ph/0702097}{{\shilacred \ttfamily hep-ph/0702097 [HEP-PH]}}].

\bibitem{Glashow:1976nt}
  S.~L.~Glashow and S.~Weinberg,
  ``Natural Conservation Laws for Neutral Currents,''
  \href{https://doi.org/10.1103/PhysRevD.15.1958}{ \tcmag{Phys.\ Rev.\ D {\bf 15} (1977) 1958}}.
  
\bibitem{Paschos:1976ay}
  E.~A.~Paschos,
  ``Diagonal Neutral Currents,''
 \href{https://doi.org/10.1103/PhysRevD.15.1966}{ \tcmag{ Phys.\ Rev.\ D {\bf 15} (1977) 1966}}.
 
\bibitem{Gu:2017ckc}
  J.~Gu, H.~Li, Z.~Liu, S.~Su and W.~Su,
  ``Learning from Higgs Physics at Future Higgs Factories,''
 \href{https://doi.org/10.1007/JHEP12(2017)153}{ \tcmag{ JHEP {\bf 1712} (2017) 153}}
  [\href{https://arxiv.org/abs/1709.06103}{{\shilacred \ttfamily arXiv:1709.06103 [hep-ph]}}].

\bibitem{Bernon:2015wef}
J.~Bernon, J.~F.~Gunion, H.~E.~Haber, Y.~Jiang and S.~Kraml,
\href{https://doi.org/10.1103/PhysRevD.93.035027}{ \tcmag{Phys. Rev. D \textbf{93} (2016) no.3, 035027}}
[\href{https://arxiv.org/abs/1511.03682}{{\shilacred \ttfamily arXiv:1511.03682 [hep-ph]}}].

\bibitem{Bernon:2015qea}
J.~Bernon, J.~F.~Gunion, H.~E.~Haber, Y.~Jiang and S.~Kraml,
``Scrutinizing the alignment limit in two-Higgs-doublet models: m$_h$=125  GeV,''
\href{https://doi.org/10.1103/PhysRevD.92.075004}{ \tcmag{Phys. Rev. D \textbf{92} (2015) no.7, 075004}}
[\href{https://arxiv.org/abs/1507.00933}{{\shilacred \ttfamily arXiv:1507.00933 [hep-ph]}}].
  
\bibitem{Dorsch:2016tab}
  G.~C.~Dorsch, S.~J.~Huber, K.~Mimasu and J.~M.~No,
  ``Hierarchical versus degenerate 2HDM: The LHC run 1 legacy at the onset of run 2,''
  \href{https://doi.org/10.1103/PhysRevD.93.115033}{ \tcmag{Phys.\ Rev.\ D {\bf 93} (2016) no.11,  115033}}
  [\href{https://arxiv.org/abs/1601.04545}{{\shilacred \ttfamily arXiv:1601.04545 [hep-ph]}}].
  
\bibitem{Dev:2014yca}
  P.~S.~Bhupal Dev and A.~Pilaftsis,
  ``Maximally Symmetric Two Higgs Doublet Model with Natural Standard Model Alignment,''
 \href{https://doi.org/10.1007/JHEP12(2014)024}{ \tcmag{ JHEP {\bf 1412} (2014) 024}}
   Erratum: [\href{https://doi.org/10.1007/JHEP11(2015)147}{\tcmag{JHEP {\bf 1511} (2015) 147}}]
  [\href{https://arxiv.org/abs/1408.3405}{{\shilacred \ttfamily arXiv:1408.3405 [hep-ph]}}].
  
\bibitem{Pilaftsis:2016erj}
  A.~Pilaftsis,
  ``Symmetries for standard model alignment in multi-Higgs doublet models,''
  \href{https://doi.org/10.1103/PhysRevD.93.075012}{ \tcmag{Phys.\ Rev.\ D {\bf 93} (2016) no.7,  075012}}
  [\href{https://arxiv.org/abs/1602.02017}{{\shilacred \ttfamily arXiv:1602.02017 [hep-ph]}}].
  
\bibitem{Das:2017zrm}
  D.~Das, U.~K.~Dey and P.~B.~Pal,
  ``Quark mixing in an $S_3$ symmetric model with two Higgs doublets,''
  \href{https://doi.org/10.1103/PhysRevD.96.031701}{ \tcmag{Phys.\ Rev.\ D {\bf 96} (2017) no.3,  031701}}
  [\href{https://arxiv.org/abs/1705.07784}{{\shilacred \ttfamily arXiv:1705.07784 [hep-ph]}}].
  
\bibitem{Benakli:2018vjk}
  K.~Benakli, Y.~Chen and G.~Lafforgue-Marmet,
  ``R-symmetry for Higgs alignment without decoupling,''
  \href{https://doi.org/10.1140/epjc/s10052-019-6676-9}{\tcmag{ Eur. Phys. J. C \textbf{79}, no.2, 172 (2019)}}
  \href{https://arxiv.org/abs/1811.08435}{{\shilacred \ttfamily arXiv:1811.08435 [hep-ph]}}].
  
\bibitem{Pramanick:2017wry}
  S.~Pramanick and A.~Raychaudhuri,
  ``Three-Higgs-doublet model under A4 symmetry implies alignment,''
  \href{https://doi.org/10.1007/JHEP01(2018)011}{ \tcmag{JHEP {\bf 1801} (2018) 011}}
  [\href{https://arxiv.org/abs/1710.04433}{{\shilacred \ttfamily arXiv:1710.04433 [hep-ph]}}].

\bibitem{Craig:2013hca}
  N.~Craig, J.~Galloway and S.~Thomas,
  ``Searching for Signs of the Second Higgs Doublet,''
  [\href{https://arxiv.org/abs/1305.2424}{{\shilacred \ttfamily arXiv:1305.2424 [hep-ph]}}].
  
\bibitem{Ferreira:2014qda}
  P.~M.~Ferreira, R.~Guedes, J.~F.~Gunion, H.~E.~Haber, M.~O.~P.~Sampaio and R.~Santos,
  ``The Wrong Sign limit in the 2HDM,''
  [\href{https://arxiv.org/abs/1410.1926}{{\shilacred \ttfamily arXiv:1410.1926 [hep-ph]}}].
  
\bibitem{Modak:2016cdm}
  T.~Modak, J.~C.~Romao, S.~Sadhukhan, J.~P.~Silva and R.~Srivastava,
  ``Constraining wrong-sign $hbb$ couplings with $h \rightarrow \Upsilon \gamma$,''
  \href{https://doi.org/10.1103/PhysRevD.94.075017}{ \tcmag{ Phys.\ Rev.\ D {\bf 94} (2016) no.7,  075017}}
  [\href{https://arxiv.org/abs/1607.07876}{{\shilacred \ttfamily arXiv:1607.07876 [hep-ph]}}].
  
\bibitem{Ferreira:2014naa}
  P.~M.~Ferreira, J.~F.~Gunion, H.~E.~Haber and R.~Santos,
  ``Probing wrong-sign Yukawa couplings at the LHC and a future linear collider,''
  \href{https://doi.org/10.1103/PhysRevD.89.115003}{ \tcmag{ Phys.\ Rev.\ D {\bf 89} (2014) no.11,  115003}}
  [\href{https://arxiv.org/abs/1403.4736}{{\shilacred \ttfamily arXiv:1403.4736 [hep-ph]}}].

\bibitem{ATLAS:2017nxi}
 [ATLAS],
``Search for heavy resonances decaying to a $W$ or $Z$ boson and a Higgs boson in final states with leptons and $b$-jets in 36.1\textasciitilde{}fb$^{-1}$ of $pp$ collision data at $\sqrt s = 13$\textasciitilde{}\textbackslash{}TeV\textbackslash{} with the ATLAS detector,''
ATLAS-CONF-2017-055.
  
  
\bibitem{Sirunyan:2018iwt}
  A.~M.~Sirunyan {\it et al.} [CMS Collaboration],
  ``Search for Higgs boson pair production in the $\gamma\gamma\mathrm{b\overline{b}}$ final state in pp collisions at $\sqrt{s}=$ 13 TeV,''
  [\href{https://arxiv.org/abs/1806.00408}{{\shilacred \ttfamily arXiv:1806.00408 [hep-ex]}}].
  
\bibitem{TheATLAScollaboration:2016loc}
  The ATLAS collaboration,
  ``Search for a CP-odd Higgs boson decaying to Zh in pp collisions at $\sqrt{s} = 13$ TeV with the ATLAS detector,''
  ATLAS-CONF-2016-015.
  
\bibitem{ATLAS:2017mpg}
  The ATLAS collaboration [ATLAS Collaboration],
  ``Search for additional heavy neutral Higgs and gauge bosons in the ditau final state produced in 36.1 fb$^{-1}$ of $pp$ collisions at $\sqrt{s}$ = 13 TeV with the ATLAS detector,''
  ATLAS-CONF-2017-050.
  
\bibitem{CMS:2015mca}
  CMS Collaboration [CMS Collaboration],
  ``Search for additional neutral Higgs bosons decaying to a pair of tau leptons in $pp$ collisions at $\sqrt{s}$ = 7 and 8 TeV,''
  CMS-PAS-HIG-14-029.
  
  
\bibitem{Dorsch:2014qja}
  G.~C.~Dorsch, S.~J.~Huber, K.~Mimasu and J.~M.~No,
  ``Echoes of the Electroweak Phase Transition: Discovering a second Higgs doublet through $A_0 \rightarrow ZH_0$,''
  \href{https://doi.org/10.1103/PhysRevLett.113.211802}{ \tcmag{ Phys.\ Rev.\ Lett.\  {\bf 113} (2014) no.21,  211802}}
  [\href{https://arxiv.org/abs/1405.5537}{{\shilacred \ttfamily arXiv:1405.5537 [hep-ph]}}].
  

\bibitem{Dorsch:2013wja}
  G.~C.~Dorsch, S.~J.~Huber and J.~M.~No,
  ``A strong electroweak phase transition in the 2HDM after LHC8,''
  \href{https://doi.org/10.1007/JHEP10(2013)029}{ \tcmag{ JHEP {\bf 1310} (2013) 029}}
  [\href{https://arxiv.org/abs/1305.6610}{{\shilacred \ttfamily arXiv:1305.6610 [hep-ph]}}].
 
\bibitem{Gao:2016ued}
  C.~Gao, M.~A.~Luty, M.~Mulhearn, N.~A.~Neill and Z.~Wang,
  ``Searching for Additional Higgs Bosons via Higgs Cascades,''
  \href{https://doi.org/10.1103/PhysRevD.97.075040}{ \tcmag{ Phys.\ Rev.\ D {\bf 97} (2018) no.7,  075040}}
  [\href{https://arxiv.org/abs/1604.03108}{{\shilacred \ttfamily arXiv:1604.03108 [hep-ph]}}].
  
\bibitem{Bauer:2015fxa}
  M.~Bauer, M.~Carena and K.~Gemmler,
  ``Flavor from the Electroweak Scale,''
  \href{https://doi.org/10.1007/JHEP11(2015)016}{ \tcmag{ JHEP {\bf 1511} (2015) 016}}
  [\href{https://arxiv.org/abs/1506.01719}{{\shilacred \ttfamily arXiv:1506.01719 [hep-ph]}}].
  
\bibitem{Kling:2015uba}
  F.~Kling, A.~Pyarelal and S.~Su,
  ``Light Charged Higgs Bosons to AW/HW via Top Decay,''
  \href{https://doi.org/10.1007/JHEP11(2015)051}{ \tcmag{ JHEP {\bf 1511} (2015) 051}}
  [\href{https://arxiv.org/abs/1504.06624}{{\shilacred \ttfamily arXiv:1504.06624 [hep-ph]}}].
  
\bibitem{Coleppa:2014cca}
  B.~Coleppa, F.~Kling and S.~Su,
  ``Charged Higgs search via $AW^\pm/HW^\pm$ channel,''
  \href{https://doi.org/10.1007/JHEP12(2014)148}{ \tcmag{ JHEP {\bf 1412} (2014) 148}}
  [\href{https://arxiv.org/abs/1408.4119}{{\shilacred \ttfamily arXiv:1408.4119 [hep-ph]}}].
  
  
\bibitem{Kling:2016opi}
  F.~Kling, J.~M.~No and S.~Su,
  ``Anatomy of Exotic Higgs Decays in 2HDM,''
  \href{https://doi.org/10.1007/JHEP09(2016)093}{\tcmag{JHEP {\bf 1609} (2016) 093}}
  [\href{https://arxiv.org/abs/1604.01406}{{\shilacred \ttfamily arXiv:1604.01406 [hep-ph]}}].
  
\bibitem{Adhikary:2018ise}
  A.~Adhikary, S.~Banerjee, R.~K.~Barman and B.~Bhattacherjee,
  ``Resonant heavy Higgs searches at the HL-LHC,''
  \href{https://doi.org/10.1007/JHEP09(2019)068}{\tcmag{ JHEP \textbf{09}, 068 (2019)}}
  [\href{https://arxiv.org/abs/1812.05640}{{\shilacred \ttfamily arXiv:1812.05640 [hep-ph]}}].
  
\bibitem{Kling:2018xud}
  F.~Kling, H.~Li, A.~Pyarelal, H.~Song and S.~Su,
  ``Exotic Higgs Decays in Type-II 2HDMs at the LHC and Future 100 TeV Hadron Colliders,''
  \href{https://doi.org/10.1007/JHEP06(2019)031}{\tcmag{ JHEP \textbf{06}, 031 (2019)}}
  [ arXiv:1812.01633 [hep-ph]].

\bibitem{Grzadkowski:2018ohf}
B.~Grzadkowski, H.~E.~Haber, O.~M.~Ogreid and P.~Osland,
``Heavy Higgs boson decays in the alignment limit of the 2HDM,''
\href{https://doi.org/10.1007/JHEP12(2018)056}{\tcmag{ JHEP \textbf{12} (2018), 056}}
[\href{https://arxiv.org/abs/1808.01472}{{\shilacred \ttfamily arXiv:1808.01472 [hep-ph]}}].

\end{thebibliography}
\end{document}